\documentclass{article}
\usepackage{amssymb}
\usepackage{latexsym}

\title{\large \bf The supersymmetric extension of the Faddeev model}
\author{\vspace{-0.1cm}\normalsize Lisa Freyhult\footnote{lisa.freyhult@teorfys.uu.se}\\ \vspace{-0.1cm}\normalsize \textsl{Department
of Theoretical Physics, Uppsala University}\\\normalsize \textsl{P.O. Box 803, S-75108, 
Uppsala, Sweden}} 

\date{}

\begin{document}
\maketitle
\begin{abstract}
We study the supersymmetric extension of the Faddeev model in four
dimensions. The Faddeev model contains three dimensional soliton
solutions and we are interested in how these solitons are affected by
supersymmetry. We consider both the $N=1$ and $N=2$ extensions and find
that in neither case it is possible to supersymmetrize the model without
adding additional bosonic terms. There are essentially two ways of
constructing the supersymmetric theory, one that will lead to a model
which allows for solitons and another that gives a model where solitons are excluded.

The $N=2$ model is studied
since extending supersymmetry is the natural way of including
topological charges in the algebra. A lower bound to the mass is obtained by computing
the central charge. The result is that it is possible to
have a non-trivial lower bound on the mass, this in principle allows
for massive solitons.

\end{abstract}
\section{Introduction}

The Faddeev model \cite{Antti2} is a non-linear sigma model with higher derivative
terms. It has been proposed as a low energy limit of Yang-Mills theory. The higher derivative terms makes three
dimensional topological soliton solutions possible \cite{Antti3}. Properties as a
topological charge and a lower bound to the mass is associated to the
solitons. 

We will study the supersymmetric extension of the model. Supersymmetry
might alter the properties of the solitons or even exclude them as solutions. We will consider if it is still possible to have the three
dimensional solitons found in \cite{Antti3} in a supersymmetric model
and, in that case, we will study their modified properties. One might also
be interested in finding the corresponding fermionic solutions to the
equations of motion.

We start by studying the $N=1$ supersymmetric extension to the model.
This has essentially been discussed before in the context of
supersymmetric skyrmeons \cite{Bergshoeff}. We review the results
that we need here and explain how to translate the results in the context of
the Faddeev model. We find that it is not possible to supersymmetrize the
model as it stands, that is without adding any new bosonic terms to it.
Considering solitons this could mean that the stability of the solutions is affected.
We will essentially discuss two ways of writing the supersymmetric
model. The action will contain fields that
belong to the vector and hypermultiplet. Writing the action we will have
the freedom of choosing the relative sign between the actions of the two
multiplets. One of the choices will lead to a model where solitons are allowed, the other results in a model were they are excluded.

We then consider the $N=2$ supersymmetric model. This is motivated by
that the central charge should give us a bound to the mass of our
solutions. We also expect the eventual solutions to have nontrivial
topological charge and extending supersymmetry is the natural way to
include that charge in the algebra. New terms still has to be added to
the original model in order to have supersymmetry, we discuss the
options in the $N=2$ case. We compute the
central charge and find that it will depend both on the fermionic and
bosonic fields. We consider the bosonic part of the central charge,
which is relevant for bosonic solitons, and find a lower bound for the
mass. The lower bound will depend on a topological invariant very
similar to the Hopf invariant and from that we get a hint of what is required of a solution with nontrivial mass.

\section{The supersymmetric Faddeev model}
We choose to study the Faddeev model in 3+1 dimensions, this corresponds to the low energy limit of SU(2) Yang-Mills theory \cite{Antti2}.
The Lagrangian for the Faddeev model is \cite{Antti2} \cite{Faddeev} :
\begin{equation}
L=\Lambda^2\partial_\mu n^a\partial^\mu n^a-(n^a\epsilon^{abc}\partial_\mu n^b\partial_\nu n^c)^2
\end{equation}
where $n$ is a three component unit vector. We use the metric with signature (1\, -1\, -1\, -1).
The second term is a closed two form squared and can be interpreted as
$F_{\mu\nu}^2=(\partial_\mu A_\nu-\partial_\nu A_\mu)^2$ with:
\begin{equation}
F_{\mu\nu}=n^a\epsilon^{abc}\partial_\mu n^b\partial_\nu n^c
\end{equation}
Here $A_\mu$ can not locally be expressed in terms of $n$. The new
variable $Z$ is therefore introduced. The vector $n$ is related to $Z$
as
\begin{equation}
n^a=\bar{Z} T^a Z
\end{equation}
where $Z=(Z_1,Z_2)$ and $Z_{1,2}$ are complex scalar fields. We use $T^a$ to denote
the generators of SU(2) in the fundamental representation.
In terms of $Z$ the vector field and the field strength are:
\begin{eqnarray}
&&A_\mu=\frac{i}{2}(\bar{Z}\partial_\mu Z-\partial_\mu \bar{Z} Z)\\
&&F_{\mu\nu}=i(\partial_\mu \bar{Z}\partial_\nu Z-\partial_\nu \bar{Z}\partial_\mu Z)
\end{eqnarray}
The condition that $n$ is a unit vector is translated into 
\begin{equation}
\bar{Z} Z=1
\end{equation}
In terms of $Z$ the Lagrangian is written as:
\begin{equation}
L=\Lambda^2\bar{D}_\mu \bar{Z} D^\mu Z-(\partial_\mu \bar{Z}\partial_\nu Z-\partial_\nu \bar{Z}\partial_\mu Z)^2
\end{equation}
with $D_\mu=\partial_\mu+iA_\mu$. Note that when introducing the
variable $Z$ instead of $n$ we have introduced an extra degree of
freedom, we could eliminate that degree of freedom by fixing the gauge of $A_\mu$.

Using $Z$ instead of $n$ is also more appropriate as $Z$ can be considered as one of the fields in the chiral multiplet. The chiral
multiplet contains a complex scalar.
Under supersymmetry the fields in the covariantly chiral multiplet transform in the following way:
\begin{eqnarray}
&&\delta Z^I=\sqrt 2\xi^\alpha\psi^I_\alpha\\
&&\delta \psi^I_\alpha=i\sqrt 2\sigma^\mu_{\alpha\dot{\alpha}}\bar{\xi}^{\dot{\alpha}}D_\mu Z^I+\sqrt 2\xi_\alpha F^I\\
&&\delta F^I=i\sqrt 2\bar{\xi}_{\dot{\alpha}}\bar{\sigma}^{\mu\dot{\alpha}\alpha}D_\mu\psi^I_\alpha-2i\bar{\xi}_{\dot{\alpha}}Z^I\bar{\lambda}^{\dot{\alpha}}
\end{eqnarray}
We have used capital latin letters to denote the components of $Z$, i.e. $I=1,2$.
The vector field $A_\mu$ depends on the scalar field $Z$ but to start
with we consider them as independent variables. We then expect $A_\mu$
to transform as part of the vector multiplet, the variations of the
fields under supersymmetry are
\begin{eqnarray}
&&\delta
A_\mu=
i(\bar{\xi}_{\dot{\alpha}}\bar{\sigma}_\mu^{\dot{\alpha}\alpha}
\lambda_\alpha+\xi^\alpha\sigma_{\mu\alpha\dot{\alpha}}\bar{\lambda}^{\dot{\alpha}})\\
&&\delta \lambda_\alpha=\frac{1}{2}\xi^\beta\sigma^{\mu\nu}_{\beta\alpha}F_{\mu\nu}+i\xi_\alpha D\\
&&\delta
D=
\bar{\xi}_{\dot{\alpha}}\bar{\sigma}^{\mu\dot{\alpha}\alpha}\partial_\mu\lambda_\alpha-\xi^\alpha\sigma^\mu_{\alpha\dot{\alpha}}\partial_\mu\bar{\lambda}^{\dot{\alpha}}
\end{eqnarray}
In order to find the supersymmetric theory we start by generalising the condition (6)  by varying it with respect to supersymmetry. 
The following constraints are obtained.
\begin{eqnarray} 
&&Z^I\bar{Z}_I =1 \\
&& Z^I\bar{\psi}_I=0 \\
&& Z^I\bar{F}_I=0\\
&&A_\mu=\frac{i}{2}(\partial_\mu \bar{Z}^I Z_I-\bar{Z}^I\partial_\mu Z_I)-\frac{1}{2}\psi^{I\alpha}\sigma_{\mu\alpha\dot{\alpha}}\bar{\psi}^{\dot{\alpha}}_I\\
&&\lambda_\alpha=-\frac{1}{\sqrt 2}i\bar{F}^I\psi_{I\alpha}+\frac{1}{\sqrt2}\sigma^\mu_{\alpha\dot{\alpha}}D_\mu Z^I\bar{\psi}^{\dot{\alpha}}_I\\
&&D=\bar{D}^\mu \bar{Z}^I D_\mu
Z_I+\frac{i}{2}(\psi^{I\alpha}\sigma^\mu_{\alpha\dot{\alpha}}\bar{D}_\mu\bar{\psi}^{\dot{\alpha}}_I-D^\mu\psi^{I\alpha}\sigma_{\mu\alpha\dot{\alpha}}\bar{\psi}^{\dot{\alpha}}_I
)-\bar{F}^I F_I
\end{eqnarray}
We see that the expression for $A_\mu$ in terms of $Z$, given in (4), follows from varying the constraint on $Z$. We
could of course also have done it the other way around, imposing (4) and deriving the other expressions using the supersymmetric transformations.
A fermionic term then has to be added to (4) for the vector field to satisfy the appropriate algebra.

How to supersymmetrize the action above is in principle well known. The
first part is similar to the nonlinear sigma model with gauge interactions and for the
second part we write down the invariant action for the vector multiplet.
Since these two parts of the action are supersymmetric independently of
each other we are free to choose their relative sign. Hence we
have
\begin{eqnarray}
\nonumber S=\Lambda^2\int d^4x\big(\bar{D}^\mu \bar{Z} D_\mu
Z+i\psi\sigma^\mu
\bar{D}_\mu\bar{\psi}-iD^\mu\psi\sigma_\mu\bar{\psi}-\bar{F}
F+i\sqrt 2\bar{Z}\lambda^\alpha\psi_\alpha\\
-i\sqrt 2Z\bar{\lambda}_{\dot{\alpha}}\bar{\psi}^{\dot{\alpha}}+D(\bar{Z}
Z-1)\big)\mp\int d^4x\big(
F_{\mu\nu}^2+
2i\lambda\sigma^\mu\partial_\mu\bar{\lambda}-
2i\partial^\mu\lambda\sigma_\mu\bar{\lambda}-2D^2\big)
\end{eqnarray}
If we now substitute for the fields in (17)-(19) we obtain the action in
terms of the relevant fields. Note that when we do this we get, from the
term that contains derivatives of the fermions in the vector multiplet,
terms that contain derivatives of the field $F$. This means that the,
previously, auxiliary fields become dynamical. Hence in order to make
the model supersymmetric we have to add, additional to the fermions,
another type of bosonic field. This is expected since in the original
action (7) we have a term with more than two derivatives.

In the above we have introduced the freedom to choose the sign in front
of the second part of the action. Comparing the supersymmetric action to
the action of the bosonic model we choose the minus sign. This directly reproduces the terms in the bosonic action.
The action in terms of the field $Z$ that is needed in order to
construct the supersymmetric model is then
\begin{equation}
S=\int d^4x\left(\Lambda^2\bar{D}^\mu \bar{Z} D_\mu Z-F_{\mu\nu}^2+2(\bar{D}^\mu \bar{Z} D_\mu Z)^2\right)
\end{equation}
The last term comes from the $D^2$ in (20).
The conclusion is that the Faddeev model can not be supersymmetrized as
it stands, another term has to be added. This term is fourth
order in time derivatives and the model therefore lacks an hamiltonian interpretation.

What is more is that when considering solitons in this model one will find
that the new term acts unstabilizing. This is seen by considering the energy of a
static field configuration and applying Derrick's theorem
\cite{Derrick} to it. 
To make this a little more transparent we go back to the original variables, $n$. The action (21) is then 
\begin{equation}
S=\int d^4x\left(\Lambda^2\partial^\mu n\partial_\mu n-(\partial^\mu n\times\partial^\nu n)(\partial_\mu n\times\partial_\nu n)+2(\partial^\mu n\partial_\mu n)^2\right)
\end{equation}
Since we are interested in how the new term affects solitons in the
model we consider the energy for static fields 
\begin{eqnarray}
E&=&\int d^3x\left(\Lambda^2\partial_i n\partial_i n+F_{ij}^2-2(\partial_i n\partial_i n)^2\right)\\
\nonumber&=&\int d^3x\left(\Lambda^2\partial_i n\partial_i n+(\partial_i n\partial_i n)^2-(\partial_i n\partial_j n)^2-2(\partial_i n\partial_i n)^2\right)\\
&=&\int d^3x\left(\Lambda^2\partial_i n\partial_i n-(\partial_i n\partial_i n)^2-(\partial_i n\partial_j n)^2\right)
\end{eqnarray}
We now apply Derrick's theorem to (24).
One finds that as $x\to\lambda x$ the term proportional to $\Lambda^2$, we call it $E_2$,
scales as $\lambda$ and the other terms, we call them $E_4$ scales as $1/\lambda$. If $\lambda=1$ corresponds to a minimum of the energy we have that
\begin{equation}
0=\frac{d E}{d\lambda}\Big|_{\lambda=1}=\frac{d }{d\lambda}(\lambda E_2-\frac{1}{\lambda}E_4)\Big|_{\lambda=1}=E_2+E_4
\end{equation}
Both $E_2$ and $E_4$ are positive definite and hence they both have to be zero for the above to hold.
Hence soliton solutions are not possible in the model.

There is a possibility to restore the loss of stability in the model by adding other
terms that are also fourth order in derivatives and at the same time
keeping supersymmetry, see \cite{Bergshoeff}. In terms of $Z$ the terms
that we could add would be $-(\bar{D}\bar{Z}DZ)^2+\Box \bar{Z}\Box Z$ times some arbitrary constant.
We will not pursue this direction and refer to \cite{Bergshoeff} for
further details.
We will instead go back to the action (20) and choose the other
possibility for the sign in front of the second part of the action. The action in terms of the original variable, $Z$, is then 
\begin{equation}
S=\int d^4x\left(\Lambda^2\bar{D}^\mu \bar{Z} D_\mu Z+F_{\mu\nu}^2-2(\bar{D}^\mu \bar{Z} D_\mu Z)^2\right)
\end{equation}
The corresponding energy in terms of $n$ is
\begin{eqnarray}
&&E=\int d^3x\left(\Lambda^2\partial_i n\partial_i n+(\partial_i n\partial_i n)^2+(\partial_i n\partial_j n)^2\right)\\
&&=\int d^3x\left(\Lambda^2\partial_i n\partial_i n+F_{ij}^2+2(\partial_i n\partial_j n)^2\right)
\end{eqnarray}
Hence we see that this is an alternative way to write the extension of the Faddeev model.
Applying Derrick's theorem to this expression solitons are not ruled
out.

Summarizing we can not obtain a supersymmetric version of the Faddeev
model. We have to add terms that are fourth order in time derivatives.
Both (21) and (26) are models that can be supersymmetrized and they both
contain the action of the Faddeev model plus a term that is fourth
order in derivatives. In (26) solitons are possible while in (21)
they are not. 

The energy of the full model corresponding to (26) is
\begin{eqnarray}
\nonumber E=\Lambda^2\int d^4x\big(\bar{D}_i \bar{Z} D_i
Z+i\psi\sigma_i
\bar{D}_i\bar{\psi}-iD_i\psi\sigma_i\bar{\psi}+\bar{F}
F-i\sqrt 2\bar{Z}\lambda^\alpha\psi_\alpha\\
+i\sqrt 2Z\bar{\lambda}_{\dot{\alpha}}\bar{\psi}^{\dot{\alpha}}-D(\bar{Z}
Z-1)\big)+\int d^4x\big(
-F_{ij}^2+
2i\lambda\sigma_i\partial_i\bar{\lambda}-
2i\partial_i\lambda\sigma_i\bar{\lambda}+2D^2\big)
\end{eqnarray}
With
\begin{eqnarray}
&&A_i=\frac{i}{2}(\partial_i \bar{Z} Z-\bar{Z}\partial_i Z)-\frac{1}{2}\psi^{\alpha}\sigma_{i\alpha\dot{\alpha}}\bar{\psi}^{\dot{\alpha}}\\
&&\lambda_\alpha=-\frac{1}{\sqrt 2}i\bar{F}\psi_{\alpha}-\frac{1}{\sqrt2}\sigma_{i\alpha\dot{\alpha}}D_i Z\bar{\psi}^{\dot{\alpha}}\\
&&D=-\bar{D}_i \bar{Z} D_i
Z-\frac{i}{2}(\psi^{\alpha}\sigma_{i\alpha\dot{\alpha}}\bar{D}_i\bar{\psi}^{\dot{\alpha}}+D_i\psi^{\alpha}\sigma_{i\alpha\dot{\alpha}}\bar{\psi}^{\dot{\alpha}}
)-\bar{F} F
\end{eqnarray}
We obtain a lower bound to the energy of the model as
\begin{eqnarray}
&&E\geq\int d^3x\left(\Lambda^2\partial_i n\partial_i n+F_{ij}^2+2(\partial_i n\partial_j n)^2\right)\\
&&\geq\int d^3x\left(\Lambda^2\partial_i n\partial_i n+F_{ij}^2\right)\geq KQ_H^{3/4}
\end{eqnarray}
The last inequality we get following \cite{Vakulenko} and \cite{Ward}.
Here $Q_H$ is the Hopf invariant and $K$ is a constant. The Hopf invariant is given by
\begin{equation}
Q_H=\frac{1}{(8\pi)^2}\int d^3x \epsilon^{ijk}F_{ij}A_k
\end{equation}
Hence if the solitons have a nontrivial Hopf invariant their mass is nonzero.
Note that this is the same bound to the mass as in the non-supersymmetric
model \cite{Vakulenko} \cite{Ward}. It is possible that the other terms affects the bound so that it
could be higher in the supersymmetric model. However it is not easy to
say because of the form of the terms. Here we are just interested in
finding some bound which can give a nontrivial mass, for that the above
is enough.

\section{N=2 supersymmetric model}
The proper way of introducing topological charges in supersymmetric
theories is to consider extended supersymmetry \cite{Witten}. We saw
that the bosonic part of the $N=1$ model, as well as the original bosonic
model, gave a bound to the mass that was proportional to the Hopf
invariant to the power $3/4$. In extended supersymmetry 
a lower bound to the mass is proportional to the central charge in
the model \cite{Witten}. Hence we expect a mass bound that is linear in
the topological invariant for the $N=2$ model, something that seems to
be very different from what we had in the $N=1$ model. We do not expect
a linear bound from the Faddeev model and it is therefore interesting to
study this case where it seems that we get just that.
We therefore proceed and construct the $N=2$ version of the Faddeev
model. 

The variations of the fields in the $N=2$ hypermultiplet coupled to the vector multiplet are
\begin{eqnarray}
&&\delta Z^I=\sqrt2\xi^{\alpha I}\psi_\alpha\\
&&\delta\psi_\alpha=i\sqrt2\sigma^\mu_{\alpha\dot{\alpha}}\bar{\xi}^{\dot{\alpha}I}D_\mu Z_I+i\sqrt2\phi\xi_{\alpha}^I\bar{Z}_I+\sqrt2\xi_{\alpha}^IF_I\\
&&\delta F^I=i\sqrt2\bar{\xi}_{\dot{\alpha}}^I\bar{\sigma}^{\mu\dot{\alpha}\alpha}D_\mu\psi_\alpha-i\sqrt2\bar{\xi}_{\dot{\alpha}}^I\phi\bar{\psi}^{\dot{\alpha}}-2i\bar{\xi}_{\dot{\alpha}}^I\bar{\lambda}^{\dot{\alpha K}}Z_K\\
&&\delta A_\mu=i(\bar{\xi}_{\dot{\alpha}}^I\bar{\sigma}_\mu^{\dot{\alpha}\alpha}\lambda_{I\alpha}+\xi^{\alpha I}\sigma_{\mu\alpha\dot{\alpha}}\bar{\lambda}_I^{\dot{\alpha}})\\
&&\delta \phi=2\xi^{\alpha I}\lambda_{I\alpha}\\
&&\delta\lambda_{\alpha}^I=\frac{1}{2}\xi^{\beta I}\sigma^{\mu\nu}_{\beta\alpha}F_{\mu\nu}+i\sigma^\mu_{\alpha\dot{\alpha}}\partial_\mu\phi\bar{\xi}^{\dot{\alpha}I}+i\xi_{\alpha }^ID\\
&&\delta D=\bar{\xi}_{\dot{\alpha}}^I\bar{\sigma}^{\mu\dot{\alpha}\alpha}\partial_\mu\lambda_{I\alpha}-\xi^{\alpha I}\sigma^\mu_{\alpha\dot{\alpha}}\partial_\mu\bar{\lambda}^{\dot{\alpha}}_I
\end{eqnarray}
Varying the constraint (6) under this new supersymmetry we obtain
expressions for the fields in the vector multiplet in terms of the fields
in the hypermultiplet 
\begin{eqnarray}
&&\bar{Z}^IZ_I=1\\
&&\bar{\xi}^I_{\dot{\alpha}}\bar{\psi}^{\dot{\alpha}}Z_I+\xi^{I\alpha}\psi_{\alpha}\bar{Z}_I=0\\
&&A_\mu=\frac{i}{2}(\partial_\mu \bar{Z}^I Z_I-\bar{Z}^I\partial_\mu Z_I)-\frac{1}{2}\psi^{\alpha}\sigma_{\mu\alpha\dot{\alpha}}\bar{\psi}^{\dot{\alpha}}\\
&&\phi=-i\bar{F}^IZ_I\\
&&\lambda_{\alpha}^I=-\frac{i}{\sqrt2}\psi_\alpha \bar{F}^I-\frac{1}{\sqrt2}Z^I\sigma^\mu_{\alpha\dot{\alpha}}\bar{D}_\mu\bar{\psi}^{\dot{\alpha}}-Z^I\phi^\dagger\psi_\alpha\\
&&\nonumber D=\bar{D}^\mu \bar{Z}^I D_\mu
Z_I+\frac{i}{2}(\psi^{\alpha}\sigma^\mu_{\alpha\dot{\alpha}}\bar{D}_\mu\bar{\psi}^{\dot{\alpha}}-D^\mu\psi^{\alpha}\sigma_{\mu\alpha\dot{\alpha}}\bar{\psi}^{\dot{\alpha}}
)-\bar{F}^IF_I\\
&&-\frac{i}{2}\psi^\alpha\phi^\dagger\psi_\alpha+\frac{i}{2}\bar{\psi}_{\dot{\alpha}}\phi\bar{\psi}^{\dot{\alpha}}+\bar{Z}^I\phi^\dagger\phi Z_I+\sqrt2i\bar{Z}^I\lambda_I\psi-\sqrt2iZ^I\bar{\lambda}_I\bar{\psi}
\end{eqnarray}
The action of the $N=2$ model with a hypermultiplet in interaction with the vector multiplet is
\begin{eqnarray}
&&S=\nonumber{\Lambda}^2\int d^4x\Big(\bar{D}^\mu \bar{Z}D_\mu Z+\frac{i}{2}\psi\sigma^\mu\bar{D}_\mu\bar{\psi}-\frac{i}{2}D^\mu\psi\sigma_\mu\bar{\psi}-\bar{F} F+\sqrt2i\bar{Z}\lambda\psi\\
&&\nonumber-\sqrt2iZ\bar{\lambda}\bar{\psi}-\frac{i}{2}\psi^\alpha\phi^\dagger\psi_\alpha+\frac{i}{2}\bar{\psi}_{\dot{\alpha}}\phi\bar{\psi}^{\dot{\alpha}}+\bar{Z}\phi^\dagger\phi Z+D(\bar{Z} Z-1)\Big)\\
&&\pm4\int d^4x\Big(\partial^\mu\phi^\dagger \partial_\mu\phi
-\frac{1}{4}F_{\mu\nu}F^{\mu\nu}+\frac{i}{2}\partial^\mu\lambda\sigma_\mu\bar{\lambda}-\frac{i}{2}\lambda\sigma^\mu\partial_\mu\bar{\lambda}+\frac{D^2}{2}\Big)
\end{eqnarray}
Note that the action is supersymmetric independently of the sign in
front of the terms in the action that contains the vectormultiplet. We
saw in the $N=1$ case that the stability of solitons in the model was
dependent on exactly that sign. 

Using the expressions (43)-(48) the Lagrangian can be written in terms of
the fields in the hypermultiplet. 
The action that is needed to construct the $N=2$ model is
\begin{equation}
S=\int d^4x\left({\Lambda}^2\bar{D}^\mu \bar{Z} D_\mu Z\mp F^{\mu\nu}F_{\mu\nu}\pm 2(\bar{D}^\mu \bar{Z} D_\mu Z)^2\right)
\end{equation}
The last term comes from the $D^2$-term in (49). This is the same action that was needed in order to construct the $N=1$ model.
It seems that, analogous to the $N=1$ model, by
choosing the appropriate sign we can have stable solitons in the model.
Note however that there are other terms in the Lagrangian now that can
make the model unstable. We want to study how the solitons in
the original model are affected and we therefore proceed and consider bosonic solitons in the
supersymmetric model. We restrict our considerations to bosonic solitons
for simplicity. The previously auxiliary field, $F$, is now a dynamical
field which means that we have more than one type of bosons to take into
account. Some of the terms that contain $F$ are negative energy terms which is the reason for our concern.

The full bosonic action is
\begin{eqnarray}
&&\nonumber S_{bosonic}={\Lambda}^2\int d^4x\left(\bar{D}^\mu\bar{Z}D_\mu
Z-\bar{F}F+|\bar{Z}F|^2\right)\pm4\int d^4x\Big(-\frac{1}{4}F^{\mu\nu}F_{\mu\nu}
\\
&&+\partial^\mu(\bar{Z}
F)\partial_\mu(Z\bar{F})+\frac{1}{2}\left(\bar{D}^\mu\bar{Z}D_\mu Z-\bar{F}F+|\bar{Z}F|^2\right)^2
\Big)
\end{eqnarray}
The corresponding energy is
\begin{eqnarray}
&&\nonumber E_{bosonic}={\Lambda}^2\int d^3x\left(\bar{D}_i\bar{Z}D_i
Z+\bar{F}F-|\bar{Z}F|^2\right)+4\int d^3x\Big(\pm\frac{1}{4}F_{ij}F_{ij}
\\
&&\pm\partial_i(\bar{Z}
F)\partial_i(Z\bar{F})\mp\frac{1}{2}\left(-\bar{D}_i\bar{Z}D_i Z-\bar{F}F+|\bar{Z}F|^2\right)^2
\Big)
\end{eqnarray}
We will choose the sign such that the original bosonic energy, that we
need in order to be able to have supersymmetry, allows for soliton solutions. We
do this because we do not want to exclude the possibility of having
solitons in terms of the field $Z$ only. Considering the full bosonic
theory we are interested in how the interactions with the new bosonic
field affects the stability and mass bound of the solitons. It is
difficult to find a mass bound directly from the energy (52) since one
cannot directly determine the relative size of the terms containing $F$. However in an extended supersymmetric theory we know that we
can compute the central charge and from that obtain a mass bound
\cite{Witten}. This means that there is an indirect way to find the mass bound, we do not have to make an estimate from (52).

\section{A mass bound}

We now proceed with an attempt to compute the central charge of the
$N=2$ algebra above. The computation is well known, see for example
\cite{Alvarez-Gaume}, in the case where the fields in the
vector multiplet are independent of those in the scalar multiplet. Here
however we also have to take (43)-(48) into account.
Note that in the following we have made the choice of a negative sign in
front of the second part of the action, corresponding to the action of the
vector multiplet.

The generators of supersymmetry are
\begin{eqnarray}
&&Q_1=\nonumber\int{d^3x}\,(-4)\left(\sigma^\mu\bar{\lambda}_1(iF_{0\mu}+\tilde{F}_{0\mu})+\sigma^\mu\bar{\sigma}^0\lambda_2\partial_\mu\phi^\dagger+\sigma^0\bar{\lambda}_1D\right)\\
&&+\sqrt2\Lambda^2\left(\sigma^\mu\bar{\sigma}^0\psi\bar{D}_\mu\bar{Z}_1+\sigma^0\bar{\psi}F_2+\sigma^0Z_1\phi^\dagger\bar{\psi}+\frac{1}{\sqrt2}\bar{\sigma}^0\bar{\lambda}_1\bar{Z}Z\right)\\
&&Q_2=\nonumber\int d^3x\,(-4)\left(\sigma^\mu\bar{\lambda}_2(iF_{0\mu}+\tilde{F}_{0\mu})-\sigma^\mu\bar{\sigma}^0\lambda_1\partial_\mu\phi^\dagger+\sigma^0\bar{\lambda}_2D\right)\\
&&+\sqrt2\Lambda^2\left(\sigma^\mu\bar{\sigma}^0\psi\bar{D}_\mu\bar{Z}_2-\sigma^0\bar{\psi}F_1+\sigma^0Z_2\phi^\dagger\bar{\psi}+\frac{1}{\sqrt2}\bar{\sigma}^0\bar{\lambda}_2\bar{Z}Z\right)
\end{eqnarray}
together with their complex conjugates. The dual field strength tensor is given by $\tilde{F}^{\mu\nu}=\frac{1}{2}\epsilon^{\mu\nu\rho\sigma}F_{\rho\sigma}$.
The algebra of the $N=2$ model is
\begin{eqnarray}
&&\{\tilde{Q}^I_\alpha,\bar{\tilde{Q}}^J_{\dot{\beta}}\}=2i\sigma^\mu_{\alpha\dot{\beta}}P_\mu\delta^{IJ}\\
&&\{\tilde{Q}^I_\alpha,\tilde{Q}^J_\beta\}=2Z^{IJ}\epsilon_{\alpha\beta}
\end{eqnarray}
where $\tilde{Q}_1=\frac{1}{\sqrt2}(Q_1+iQ_2)$ and $\tilde{Q}_2=\frac{1}{\sqrt2}(Q_1-iQ_2)$.
Here $Z^{IJ}$ is antisymmetric in its indices and called the central charge.
The central charge gives a mass bound to the eventual solitons in the
model \cite{Witten} according to
\begin{equation}
M_{SUSY}\geq \sqrt2|Z_{12}|
\end{equation}
We therefore want to compute the anticommutator in
(56). We do that using
\begin{eqnarray}
&&\{\lambda^I_\alpha,\bar{\lambda}_{J\dot{\beta}}\}=\frac{i}{4}\sigma^0_{\alpha\dot{\beta}}\delta^I_J\\
&&\{\psi_\alpha,\bar{\psi}_{\dot{\beta}}\}=\frac{i}{\Lambda^2}\sigma^0_{\alpha\dot{\beta}}
\end{eqnarray}
and the corresponding commutators for the scalar and vector field.
Assuming that we have static bosonic solitons in the model we obtain for
the central charge:
\begin{equation}
\{Q^1_\alpha,Q^2_\beta\}=8i\int d^3x\,\frac{1}{2}\epsilon^{ijk}F_{jk}\partial_i\phi^\dagger\epsilon_{\alpha\beta}\end{equation}
Hence we have that the central charge is
\begin{equation}
Z^{12}=2i\int d^3x\,\epsilon^{ijk}F_{jk}\partial_i\phi^\dagger
\end{equation}
Where $F_{ij}$ is the bosonic part of the
field strength, $F_{ij}=i(\partial_i \bar{Z}\partial_j Z-\partial_j
\bar{Z}\partial_i Z)$, and $\phi$ is given by (46). Supersymmetry gives
us the following relation for the mass of solitons in the model
\begin{equation}
M_{SUSY}\geq\sqrt2|Z^{12}|=\sqrt2|2i\int d^3x\,\epsilon^{ijk}F_{jk}\partial_i\phi^\dagger|=2\sqrt2|\int d^3x\,\epsilon^{ijk}F_{jk}\partial_i(\bar{Z}^IF_I)|
\end{equation}
In order to see if the lowest bound is non-zero we study the vacuum
configuration. We start by considering the terms in the action that
corresponds to a potential
\begin{equation}
V(Z,F)=\Lambda^2\left(F\bar{F}-(\bar{Z}F)(\bar{F}Z)\right)+2\left(F\bar{F}-(\bar{Z}F)(\bar{F}Z)\right)^2
\end{equation}
Since $F\bar{F}-(\bar{Z}F)(\bar{F}Z)\geq 0$, the potential is positive definite and we have that
the energy is minimized when $V(Z,F)=0$. This corresponds to a field
configuration where 
\begin{eqnarray}
F\bar{F}-(\bar{Z}F)(\bar{F}Z)=0\quad\Leftrightarrow\quad Z\parallel F\mbox{,}\hspace{0.3cm} |F|=|\bar{Z}F|
\end{eqnarray}
That is we have 
\begin{equation}
\bar{Z}F=Ce^{i\chi}\quad F_I=CZ_Ie^{i\chi}
\end{equation}
Where $C=\mbox{constant}$ is a special solution.
In order to compute the mass bound we introduce the parametrization of $Z$:
\begin{equation}
Z=\left(\begin{array}{c}e^{i\Phi_{12}}\sin\frac{\nu}{2}\\e^{i\Phi_{34}}\cos\frac{\nu}{2}\end{array}\right)
\end{equation}
We define $\alpha=\Phi_{12}+\Phi_{34}$, $\beta=\Phi_{34}-\Phi_{12}$
and $\gamma=\pi-\nu$ and obtain for the bosonic part of the vector field
and field strength
\begin{equation}
A=\cos\gamma d\beta+d\alpha\quad\quad\quad F=\sin\gamma d\beta\wedge d\gamma
\end{equation}
Using this we obtain for the mass bound
\begin{equation}
M_{SUSY}\geq |\sqrt2C\int\sin\gamma d\beta\wedge d\gamma\wedge d\chi|
\end{equation}
If $\gamma$, $\beta$ and $\chi$ are coordinates on $S^3$ this is, up to a constant, the winding number $\pi_3(S^3)$.
Hence if the angle $\chi$ is nontrivial we have a nontrivial bound to the mass of solutions to the model.
Note the similarity of the mass bound to the Hopf invariant
\begin{equation}
Q_H\propto\int F\wedge A=\int\sin(\gamma)d\beta\wedge d\gamma\wedge d\alpha
\end{equation}
The Hopf invariant is not directly related to the central charge in this
model we just note that the structure of the two invariants are very
similar. However the above suggest that there are possibilities to have
solitons with nontrival mass. In order to find out exactly which masses
we can have in the model we have to explicitly find the solitons.

The above suggests that the bound is linear in the topological invariant
, just as expected. However the topological invariant that we obtained
in the supersymmetric model is not necessarily the same as the Hopf
invariant in the bosonic model. In order to better understand how the
two invariants are related we will compare the two expressions when the
bounds are saturated. We begin by reviewing how the bound is obtained in
the ordinary Faddeev model.
The energy of the Faddeev model is
\begin{equation}
E=e_2+e_4
\end{equation}
where $e_2=\Lambda^2(\partial_i n)^2$ and $e_4=F_{ij}^2=(n\cdot\partial_i n\times\partial_j n)^2$.
The energy can be given a lower bound as follows
\begin{equation}
E=(\sqrt{e_2}-\sqrt{e_4})^2+2\sqrt{e_2}\sqrt{e_4}\geq2\sqrt{e_2}\sqrt{e_4}
\end{equation}
This is then related to by a Sobolev-type inequality to an expression
proportional the Hopf invariant to the power $3/4$ \cite{Vakulenko} \cite{Ward}. The bound above is
saturated when $\sqrt{e_2}=\sqrt{e_4}$. This is consistent with the result from Derrick's theorem.
The energy is then
\begin{equation}
E_{min}=2e_4
\end{equation}

We will now use the bound (62) that we obtained for the energy of the bosonic part of the supersymmetric model.
Completing the square the energy (52) can be written as
\begin{eqnarray}
&&\nonumber E_{bosonic}=\int d^3x\Big({\Lambda}^2\left(\bar{D}_i\bar{Z}D_i
Z+\bar{F}F-|\bar{Z}F|^2\right)\\
&&\nonumber-(\frac{1}{\sqrt2}\epsilon_{ijk}F_{jk}+2i\partial_i(\bar{Z}F))(\frac{1}{\sqrt2}\epsilon_{ijk}F_{jk}-2i\partial_i(Z\bar{F}))
\\&&+\sqrt2\epsilon_{ijk}F_{jk}(i\partial_i(\bar{Z}F)-i\partial_i(Z\bar{F}))+2\left(\bar{D}_i\bar{Z}D_i
Z+\bar{F}F-|\bar{Z}F|^2\right)^2\Big)
\end{eqnarray} 
We have that the energy is 
\begin{equation}
E_{bosonic}=2\sqrt2|\int d^3x\epsilon_{ijk}F_{jk}\partial_i(\bar{Z}F)|
\end{equation}
when the following equation holds
\begin{eqnarray}
\nonumber\int d^3x\Big({\Lambda}^2\left(\bar{D}_i\bar{Z}D_i
Z+\bar{F}F-|\bar{Z}F|^2\right)+2\left(\bar{D}_i\bar{Z}D_i
Z+\bar{F}F-|\bar{Z}F|^2\right)^2
\\-(\frac{1}{\sqrt2}\epsilon_{ijk}F_{jk}+2i\partial_i(\bar{Z}F))(\frac{1}{\sqrt2}\epsilon_{ijk}F_{jk}-2i\partial_i(Z\bar{F}))\Big)=0
\end{eqnarray}
One configuration that saturates this bound is
\begin{eqnarray}
&&\frac{1}{\sqrt2}\epsilon_{ijk}F_{jk}+2i\partial_i(\bar{Z}F)=0\\
&&V(Z,F)=0\\
&&\bar{D}_i\bar{Z}D_iZ=0
\end{eqnarray}
Using this the energy, when the bound is saturated, is
\begin{equation}
E_{min}=|\int d^3x\epsilon_{ijk}F_{jk}\epsilon_{ilm}F_{lm}|=2e_4
\end{equation}
This coincides with the minimal energy for the original bosonic Faddeev
model. Hence the original bound is contained in the supersymmetric
bound. Therefore, if the bound is nontrivial in the original model the
bound must also be nontrivial in the supersymmetric model. 
\section{Conclusions}
In conclusion we have found that it is not possible to supersymmetrize
the Faddeev model as it stands. We have to add new higher derivative
terms to make it possible. There are several ways to modify the model to
get a version that can be supersymmetrized. We have analyzed the
different possibilities and found that there is one version of the model
that is supersymmetric but where bosonic solitons are excluded and one
where they are allowed. We have considered both the $N=1$ and the $N=2$
versions of the model and found that the original theory have to be
modified in the same way in both cases to allow for supersymmetry. A
lower bound to the mass have been found in both models. In the $N=1$
case the bound is proportional to the Hopf invariant to the power $3/4$. In the
$N=2$ case the bound is given by an invariant very similar, but not directly
related, to the Hopf invariant. This bound is, as expected in a theory
with extended supersymmetry, linear in the topological invariant. We
have compared the saturated bounds in the different models
and found that the bound in the extended model contains
the non-supersymmetric and N=1 bound. Doing that we draw the conclusion that a nontrivial
lower
bound is possible in the model with extended supersymmetry just as in
$N=1$ and the non-supersymmetric model.

Here we have mainly considered bosonic solitons for the reason that in
the ordinary Faddeev model three dimensional solitons exist. We have
analyzed how these solitons are affected when supersymmetry is added. It
would also be of interest to find the corresponding fermionic solitons
in the model and study how the interaction between fermions and bosons
affects the solutions. This would presumably be quite involved since one
would have to solve the fermionic equations of motion in the background
of the bosonic solitons. Due to the non-linearity of the equations 
the fermionic solutions would affect the bosonic etc. What is easily
done however is to compute a bound on the mass for the full model,
containing fermions. It might from there be possible to draw some
conclusions of what is required of a fermionic soliton for it to
contribute to the mass bound. This remains for future study.
\section{Acknowledgements}
I would like to thank A.J. Niemi for helpful discussions and for
reading the manuscript. I would also like to thank S.F. Hassan, U. Lindstr\"{o}m,
N.S. Manton and P. Sundell for useful discussions.

\end{document}